# Measuring the thermal diffusivity in a student laboratory

Amelia Carolina Sparavigna
Department of Applied Science and Technology
Politecnico di Torino, Torino

The paper describes a method for measuring the thermal diffusivity of materials having a high thermal conductivity. The apparatus is rather simple and low-cost, being therefore suitable in a laboratory for undergraduate students of engineering schools, where several set-ups are often required. A recurrence numerical approach solves the thermal field in the specimen, which is depending on the thermal diffusivity of its material. The numerical method requires the temperature data from two different positions in the specimen, measured by two thermocouples connected to a temperature logger.

The evaluation of thermal properties of new materials is quite important. For several of their engineering applications in microscopic or macroscopic structures for instance, we need to know how they are able to dissipate heat. The same is true for those systems suitable for the recover or storage of energy [1]. Besides this necessity of measuring the thermal properties of new component materials, the study and development of relevant experimental methods is quite important for researchers and students of engineering schools too. Here then, we propose a method that allows the students to have an experimental approach to the problem of thermal transport.

Years ago, the author has published some papers [2-12] on new methods to measure the thermal diffusivity. Some measures reported in the references were based on modelling the thermal field inside a specimen after the measure of its thermal expansion. This dilatometric method is able to reduce strongly the experimental errors, which are derived from thermal leaks, but requires a capacitive system to record the thermal expansion. It is therefore not suitable to be used in a laboratory for undergraduate students. Moreover, the instruments are quite expensive.

Other methods, known as the flash methods, to determine the thermal diffusivity exist. They are based on the use of a laser for heating the specimen. A high-intensity short-duration light pulse is absorbed in the front surface of the specimen and the resulting temperature history of the rear surface is measured, usually by a thermocouple. To cancel the problem of the thermal leaks through the thermocouple wires, the thermal diffusivity can be measured using a non-contact experimental configuration based on infrared photothermal radiometry [15]. This technique reduces the thermal leaks, but, as in the case of dilatometric methods, requires a sophisticate set-up.

Aiming to improve a laboratory for students with a measure of the thermal diffusivity, without increasing the overalls cost of the structure, we describe here an experimental procedure based on the use of thermocouples, adapting the numerical methods of determining the thermal field developed in Ref.[2-12]. The method of solution is simplified for undergraduate students. In the following, the experimental set-up is proposed for materials having high thermal conductivity. As we will see, the method provides good results in agreement with previous measurements.

**Experimental set-up**
The instrumentation to be used is very simple. We need two thermocouples and a two-channel temperature logger. I used for students an old device having two large displays, quite useful during room demonstrations, because everybody can see directly the variations of temperature. A cylindrical specimen of commercial aluminium, 16 cm long, is placed on a metal grid, as shown in Figure 1. Around the cylinder, a layer a few millimetres thick of grey plasticine is used to avoid convective flows. The heat source is applied at the lower base. The heater is an electric resistive coil under an insulating disk. Some students simply used a lighter during the experiments.

For what concerns the specimen, to reduce the dispersion of heat from its lateral surface, a thermal guard is use of the same material (Fig.2). The temperature is measured at two different positions in the specimen. One thermocouple is placed at two millimetres from the lower base, in a small diameter hole drilled parallel to the base, through the thermal guard to the axis of the specimen. Since the wires of thermocouple are heated, as the specimen and the thermal guard are, we can consider negligible any thermal leak due to their presence. We can see from Figure 3 how the lower part of the specimen / thermal guard cylinder had been properly shaped.

The proposed set-up is suitable for measurements with materials possessing a high thermal conductivity (for low-conducting materials, an arrangement as in Ref.4 is more suitable). Due to the high thermal transport, we can assume that the thermal field does not depend on the radial coordinate. The actual length of the specimen that we will consider in the calculation of the thermal field is the distance from the drilled hole in the lower base to the upper base; this distance is equal to 15.7 cm.

The second thermocouple is placed at the top of the specimen. Of course, a computer could record the temperatures. In the case that we use a device as in Fig.1, a stopwatch allows to directly read and record the temperatures with a suitable time interval (we used an interval of 10 seconds). The data analysis is done with a numerical program, which is quite simple, as we discuss in the next section.

**Theoretical model**
An approach that we can use to solve the thermal transport in the cylindrical specimen is numerical. It is based on a recurrence relation, which uses the temperature measured by the thermocouple at the lower base. After imposing a value of the thermal diffusivity, the temperature field in the cylinder is obtained. Evaluating with the recurrence relation the field at the upper base, we can compare its time behaviour with the temperature recorded by the second thermocouple. The actual thermal diffusivity of the sample is that providing the best agreement between theoretical and experimental values.

Let us remember that the thermal diffusivity α is the physical quantity, which appears in the equation of heat:

$$\frac{1}{\alpha} \cdot \frac{\partial T}{\partial t} = \nabla^2 T \qquad (1)$$

.
To determine the thermal diffusivity, we measure the temperature as a function of time, using at least two thermocouples, as shown in the sketch (Fig.3) of the cylindrical specimen. As previously discussed, the lower base is heated by means of a suitable heat source: we assume that the lower base is uniformly heated.

Using the two thermocouples, the temperature is measured at two different positions. As previously told, one of the thermocouples is inserted laterally at the lower base, passing through the thermal guard. The wires of this thermocouple, having practically the same temperature of the specimen and of the thermal guard, have a negligible pertubative effects on the temperature field. The other thermocouple is inserted on the upper base of the specimen. The thermal guard is a hollow cylinder made of the same material of the specimen, a commercial aluminium alloy. The radial distance between the points AB and DE is of a millimetre. The outer diameter of 4 cm. That of the inner cylinder of 2 cm. The total length is 16 cm. The distance of the two thermocouples is 15.7 cm (see Fig.3).

Since the material under measurement has a high thermal conductivity, we consider its thermal field depending only on *z*, not on the radial coordinate. Considering only a dependence on the *z*-coordinate, this means that each section of the cylinder is reached in a uniform manner by the heat coming from below, and that there is not any radial propagation of heat. If the two cylinders are of the same material, positions at the same distance from the base (for example, A, B, C, D, E in Fig.3) will have the same temperature. We assume that this is true, because of the presence of the thermal guard.

Admitting that the heat spreads along the longitudinal axis *z*, and calling *T* the thermal field in the specimen:

$$\frac{\partial T}{\partial x} = \frac{\partial T}{\partial y} = 0 \qquad (2)$$

Therefore, the thermal equation (1) can be written as:

$$\frac{\partial T}{\partial t} = \alpha \frac{\partial^2 T}{\partial z^2} \qquad (3)$$

where $\alpha = K/(c\rho)$ is the thermal diffusivity, $K$ the thermal conductivity, $c$ the specific heat and $\rho$ the density. To solve this last equation it is necessary to have the thermal field at the initial time of measurements and the boundary conditions. At the initial time condition, we imagine that all the cylinder is at the same temperature, the room temperature, $T(t = 0) = T_0$.

Since the problem is one-dimensional, we need two conditions: one at the lower base, $z = 0$, and at the top of the cylinder, $z = L$, where $L$ is the length of the specimen. At the lower base, the temperature is known, because measured by the lower thermocouple. At the top of the specimen we assume a non dispersive boundary condition. From the lateral surfaces, the irradiation of heat is negligible. The thermal exchange with the environment exists for sure, but it can be neglected because air is not flowing between the sample and the thermal guard and the difference between the temperatures of specimen and environment is quite small [9]. In any case, the aim of this method is the use of it for students, and therefore, with some suitable remarks, a non-dispersive condition can be assumed.

Setting equal to zero the thermal flow between the surface and the environment, we have:

$$0 = K \frac{\partial T}{\partial z}\bigg|_L \qquad (4)$$

Let us consider a solution of Eq.3, for boundary condition (4), in the following form:

$$\sin(\omega z) \cdot \exp(-\alpha \omega^2 t) \qquad (5)$$

According to (4), we have:

$$e^{-\alpha\omega^2 t} \cdot \cos(\omega L) = 0 \rightarrow \cos(\omega L) = 0 \tag{6}$$

Therefore:

$$\omega_n L = (2n+1) \cdot \frac{\pi}{2} \tag{7}$$

where *n* is an integer ranging from 0 to infinite. Let us use the notation:

$$\phi_n(z) = \sin(\omega_n z) \tag{8}$$

The solution of the thermal transport is obtained as in Refs.[2-12]. Let us consider a time interval τ small enough. In fact, we suppose that the time is discretized according to the used data recording of the temperature:

$$\tau = t_1 - t_o = ... = t_{i+1} - t_i = t_{i+2} - t_{i+1} = ... \tag{9}$$

Assume a time $t_i$, at which temperature is recorded, and call:

$$\theta_{i-1}^i(z,t),\ \theta_i^{i+1}(z,t) \tag{10}$$

the temperature field in the specimen during the two time intervals $(t_{i-1}, t_i), (t_i, t_{i+1})$.
We can write:

$$\theta_{i-1}^i = T_{i-1}^{(1)} + \sum_{n=0}^{\infty} C_n^{i-1} \phi_n(z) \cdot e^{-\alpha\omega_n^2 t} \tag{11}$$

$$\theta_i^{i+1} = T_i^{(1)} + \sum_{n=0}^{\infty} C_n^i \phi_n(z) \cdot e^{-\alpha\omega_n^2 t} \tag{12}$$

When z = 0, θ is given by $T = T^{(1)}$, the temperature measured by the lower thermocouple. Imposing the continuity of the temperature function with respect to time, we have:

$$\theta_i^{i+1}(z,t_i) = \theta_{i-1}^i(z,t_i) \tag{13}$$

This condition gives us the possibility to evaluate the coefficients $C_n^i$ in the following manner. From the continuity condition (13), we have:

$$T_i^{(1)} + \sum_{n=0}^{\infty} C_n^i \phi_n(z) \cdot e^{-\alpha\omega_n^2 t_i} = T_{i-1}^{(1)} + \sum_{n=0}^{\infty} C_n^{i-1} \phi_n(z) \cdot e^{-\alpha\omega_n^2 t_i} \tag{14}$$

$$T_i^{(1)} - T_{i-1}^{(1)} + \sum_{n=0}^{\infty} C_n^i \phi_n(z) \cdot e^{-\alpha\omega_n^2 t_i} - \sum_{n=0}^{\infty} C_n^{i-1} \phi_n(z) \cdot e^{-\alpha\omega_n^2 t_i} = 0 \tag{15}$$

It is necessary to use the fact that:

$$T_i^{(1)} - T_{i-1}^{(1)} = \left(T_i^{(1)} - T_{i-1}^{(1)}\right) \cdot 1 = \left(T_i^{(1)} - T_{i-1}^{(1)}\right) \sum_{n=0}^{\infty} F_n \phi_n(z) \qquad (16)$$

and:

$$F_n = \frac{2}{\omega_n L} \qquad (17)$$

Then:

$$C_n^i - C_n^{i-1} = \left(-T_i^{(1)} + T_{i-1}^{(1)}\right) \cdot F_n \cdot \exp\left(\alpha \omega_n^2 t_i\right) \qquad (18)$$

The following recurrence relation can be use in the numeric calculation:

$$\begin{aligned}
C_n^0 &= 0 \\
C_n^1 - C_n^0 &= \left(T_0^{(1)} - T_1^{(1)}\right) \cdot F_n \cdot \exp\left(\alpha \omega_n^2 \tau\right) \\
C_n^2 &= C_n^1 + \left(T_1^{(1)} - T_2^{(1)}\right) \cdot F_n \cdot \exp\left(\alpha \omega_n^2 2\tau\right) = \\
&= C_n^0 + \left(T_0^{(1)} - T_1^{(1)}\right) \cdot F_n \cdot \exp\left(\alpha \omega_n^2 \tau\right) + \left(T_1^{(1)} - T_2^{(1)}\right) \cdot F_n \cdot \exp\left(\alpha \omega_n^2 2\tau\right) = \\
&= C_n^0 + \sum_{p=1}^{2} \left(T_{p-1}^{(1)} - T_p^{(1)}\right) \cdot F_n \cdot \exp\left(\alpha \omega_n^2 \tau \cdot p\right)
\end{aligned} \qquad (19)$$

For index $i$:

$$C_n^i = \sum_{p=1}^{i} \left(T_{p-1}^{(1)} - T_p^{(1)}\right) \cdot F_n \cdot \exp\left(\alpha \omega_n^2 \tau \cdot i \cdot p\right) \qquad (20)$$

In general, the temperature field of the specimen at the time $t_i = i \cdot \tau$ is therefore written as:

$$\begin{aligned}
\theta_i^{i+1} &= T_i^{(1)} + \sum_{n=0}^{\infty} C_n^i \phi_n(x) \cdot e^{-\alpha \omega_n^2 (i+1)\tau} = \\
&= T_i^{(1)} + \sum_n \sum_{p=1}^{i} \left(T_{p-1}^{(1)} - T_p^{(1)}\right) \cdot F_n \cdot \phi_n(z) \cdot e^{-\alpha \omega_n^2 \tau (i-p)}
\end{aligned} \qquad (21)$$

We can write this last equation for $z = L$, the upper base. This theoretical value is a function of the chosen thermal diffusivity.

$$\theta_i(\alpha) = \theta_i^{i+1} \tag{24}$$

We can compare it with the temperature $T^{(2)}$ recorded by the thermocouple at the upper base of the specimen, by means of the following function:

$$I(\alpha) = \sum_i \left(\theta_i(\alpha) - T_i^{(2)}\right)^2 \tag{25}$$

Minimizing (25), the value of the thermal diffusivity is obtained, giving the best agreement between the model and the experimental measured temperature.

**Discussion**
Heating the specimen for three minutes approximately and recording the two temperatures each ten seconds, two curves as in Fig.4 are obtained. Applying the numerical procedure described in the previous section, $I(\alpha)$ as a function of the thermal diffusivity is obtained. The behaviour of this function for the data of Fig.4 is given in Fig.5. This function has a well-defined minimum.
The value of the thermal diffusivity giving the best agreement with experimental data of Fig.4 is therefore 0.79 $cm^2/s$. Repeating some measurements on the same specimen, we obtained a value of the thermal diffusivity of (0.75 ± 0.07) $cm^2/s$. This measure is in good agreement with the values obtained in a more controlled environment [2,7-9,15]. The dispersion of the values is larger, but this is not surprising, because it is difficult to repeat the measures in the same conditions. Moreover, there is the possibility that convective heat flows exist during the heating of the specimen.
Of course, the method can be improved preparing a small box for the set-up and create the vacuum in it with a rotative pump, but this is beyond the aim of the proposed approach, which is the use of it in a student laboratory. For what concerns the calculation, the numerical recurrence is quite simple to be implemented by an undergraduate student on a personal computer. Moreover, the apparatus has a low-cost, and therefore several replicas can be prepared without being their cost exceedingly large for the financial support of the laboratory.

**References**
1. Sparavigna A.C., S. Giurdanella, M. Patrucco Behaviour of Thermodynamic Models with Phase Change Materials under Periodic Conditions. Energy and Power Engineering, **3**, 150-157, 2011..
http://www.scirp.org/journal/PaperInformation.aspx?paperID=4920&publishStatus=2
2. Sparavigna A., Nuovo metodo dilatometrico di misura della diffusivita' termica dei solidi, Ph:D. Dissertation, 1990.
3. Sparavigna A., Omini M., Pasquarelli A. and Strigazzi A., Thermal diffusivity of amorphous plastic materials, High Temp - High Press. **24,** 175, 1992.
4. Sparavigna A., Omini M, Pasquarelli A; Strigazzi A, Thermal diffusivity and conductivity in low-conducting materials: a new technique, Int. J. of Thermophysics, **13**, 351, 1992.
5. Omini M., Sparavigna A. and Strigazzi A., Thermal diffusivity of solids with low expansion coefficient: a dilatometric technique, Int. J. of Thermophysics, **13**, 711, 1992.
6. Giachello G., Omini M., Sparavigna A. and Strigazzi A., Thermal diffusivity in low conducting solids: a capacitive method, Physica Scripta, **42**, 439, 1990.
7. Sparavigna A., Giachello G., Omini M. and Strigazzi A., High- sensitivity capacitance method for measuring thermal diffusivity and thermal expansion: results on aluminum and copper, Int. J. of Thermophysics, **11**, 1111, 1990.
8. Omini M, Sparavigna A. and Strigazzi A, Thermal diffusivity and Biot number: a new experimental technique, Appl. Phys. A., **50**, 35, 1990.
9. Omini M., Sparavigna A., Strigazzi A. and Teppati G., A ring-source method to measure the thermal diffusivity of a conducting specimen in forced air flow: study on aluminum, Nuovo Cimento D, **12**, 79,1990.
10. Omini M., Sparavigna A. and Strigazzi A., Dilatometric determination of thermal diffusivity in low conducting materials, Measurement Science & Technology, **1**, 166, 1990.
11. Omini M., Sparavigna A., and Strigazzi A., Calibration of capacitive cells for dilatometric measurements of thermal diffusivity, Measurement Science & Technology, **1**, 1228, 1990.
12. Omini M., Sparavigna A. and Strigazzi A., Thermal diffusivity and thermal expansion in high-conducting materials: a capacitance method, Europhysics Letters, **10**, 129, 1989.
13. Parker W.J., Jenkins R.J., Butler, C.P. and Abbott, G.L., Flash Method of Determining Thermal Diffusivity, Heat Capacity, and Thermal Conductivity, Journal of Applied Physics, **32**, 1679, 1961.
14. Cape J.A. and Lehman G. W., Temperature and Finite Pulse-Time Effects in the Flash Method for Measuring Thermal Diffusivity, Journal of Applied Physics, **34**, 1909, 1963.
15. E. MacCormack, A. Mandelis, M. Munidasa, B. Farahbakhsh and H. Sang, Measurements of the Thermal Diffusivity of Aluminum Using Frequency-Scanned, Transient, and Rate Window Photothermal Radiometry. Theory and Experiment, Int. J. of Thermophysics, **18**, 221, 1997.

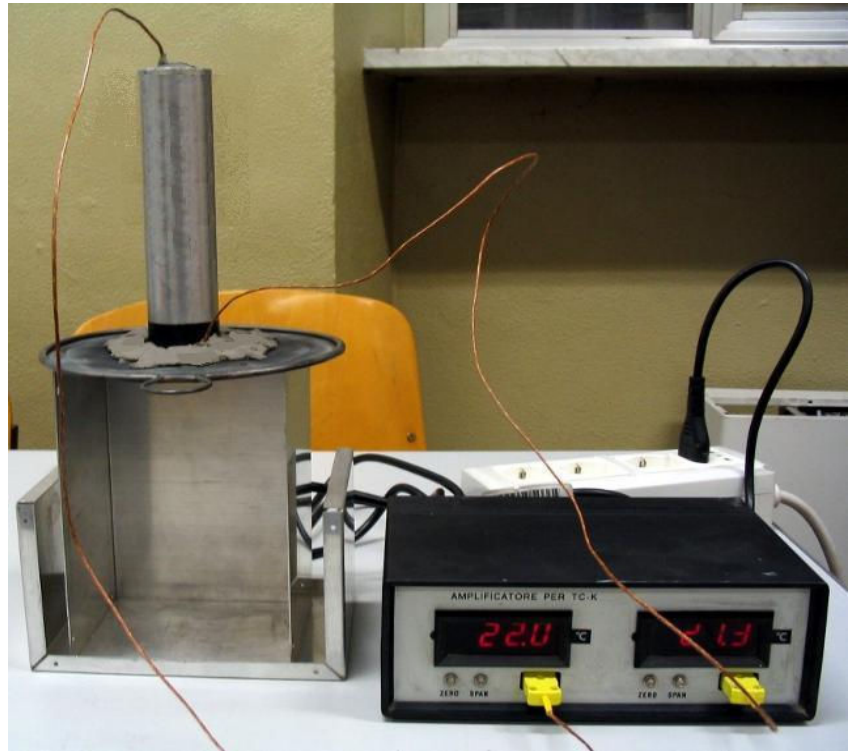

Fig.1. Experimental set-up for measuring the thermal diffusivity. We see the cylinder on the left, which is composed by the specimen and its thermal guard, placed on a metallic grid. Near the base of the cylinder, a small amount of insulating material (grey plasticine) is used to avoid convective currents about the cylinder, when the specimen is heated at its lower base. We can simply use a lighter to heat it. On the right we see the data logger for thermocouples, having two displays to visualize the temperatures. We can also see the wires of thermocouples..

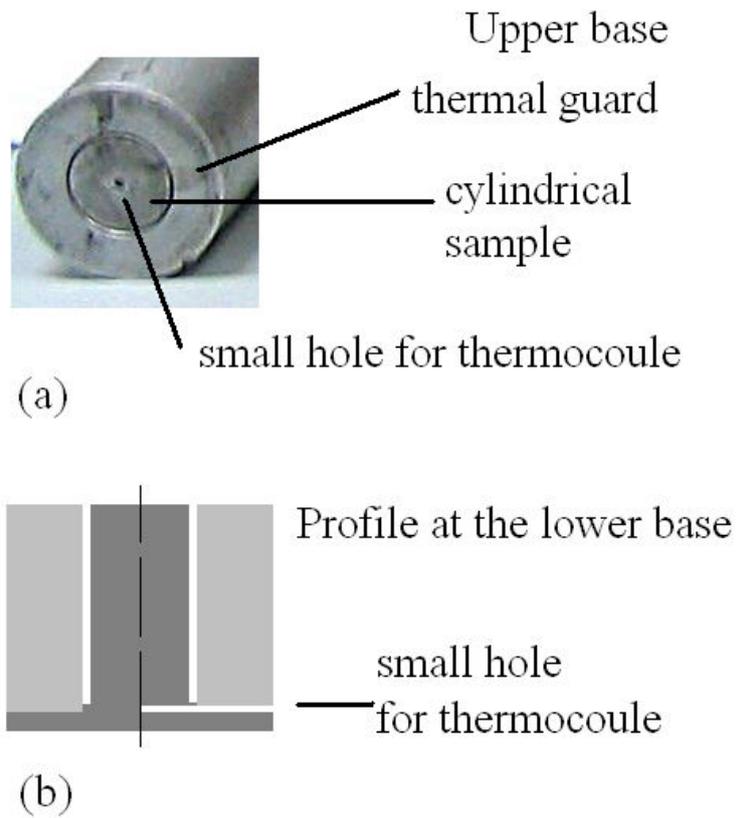

Fig.2. Panel (a) shows the upper base of cylinder and thermal guard. We can see a very small hole for the thermocouple. Panel (b) shows the profile of the specimen (dark grey) and of the thermal guard at the lower base of the cylinder. We see also the hole for the thermocouple, drilled in the cylinder.

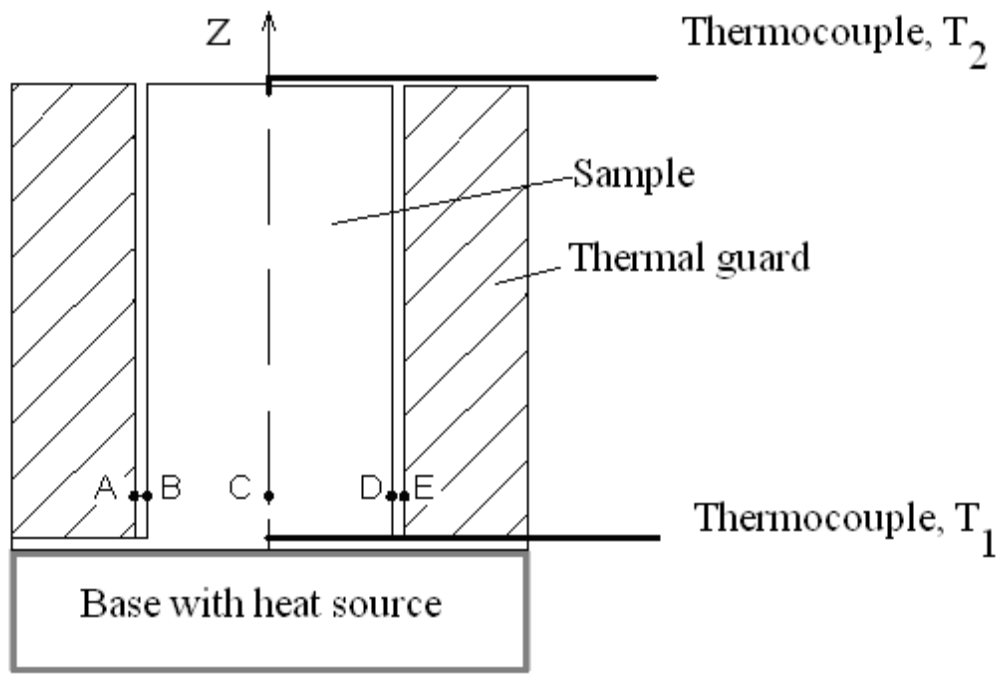

Fig.3. Sketch of specimen and thermal guard.

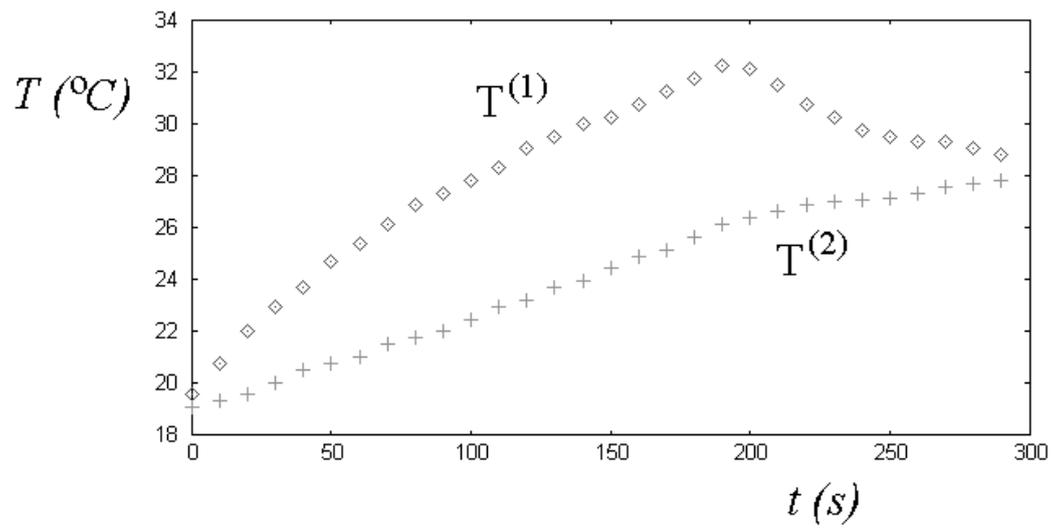

Fig.4 Temperatures recorded by the two thermocouples.

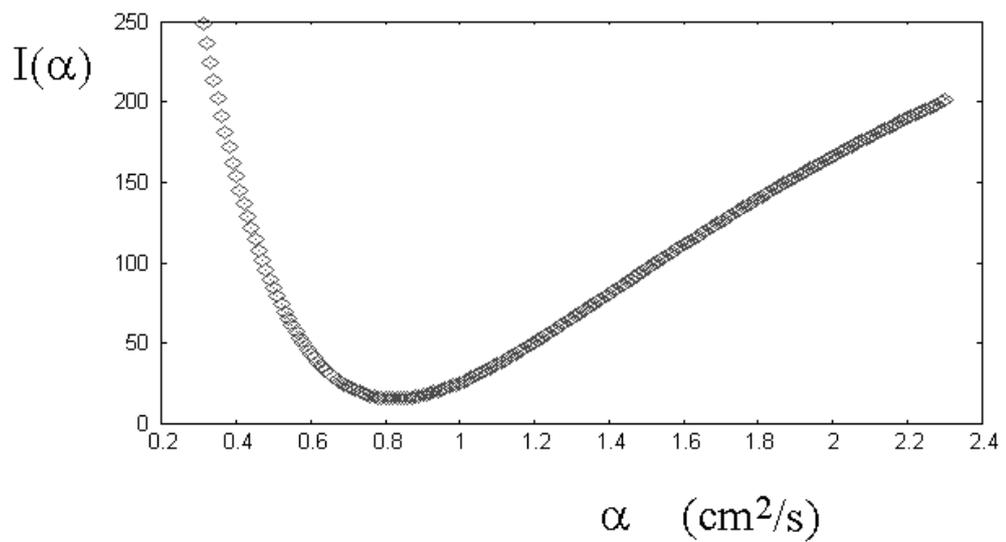

Fig.5 $I(\alpha)$ as a function of the thermal diffusivity.